\documentclass[]{spie}  %>>> use for US letter paper
\usepackage[latin1]{inputenc}
\usepackage[british]{babel}
\usepackage[]{graphicx}
\usepackage{amsmath,amssymb}
\usepackage{textcmds}
\usepackage{astro_bib_macro} %raccourcis noms de journaux d?finis dans ce fichier
\usepackage{multirow}
\usepackage{color}

\usepackage{hyperref}

\title{High-contrast imager for Complex Aperture Telescopes\\ (HiCAT): 3. first lab results with wavefront control.} 

\author{
Mamadou N'Diaye\supit{a}*, 
Johan Mazoyer\supit{a},
Elodie Choquet\supit{a},
Laurent Pueyo\supit{a},\\
Marshall D. Perrin\supit{a},
Sylvain Egron\supit{b,a},
Lucie Leboulleux\supit{b,a},
Olivier Levecq\supit{b,a},\\
Alexis Carlotti\supit{a},
Chris A. Long\supit{a},
Rachel Lajoie\supit{a},
and R\'emi Soummer\supit{a\dag}
\skiplinehalf
\supit{a} Space Telescope Science Institute, 3700 San Martin Drive, Baltimore, MD 21218, USA\\
\supit{b} Institut d'Optique Graduate School (Palaiseau, Saint-Etienne, Bordeaux), France\\
}

\authorinfo{*E-mail: mamadou@stsci.edu, \dag: HiCAT bench Principal Investigator}
\begin{document} 
  \maketitle 

%%%%%%%%%%%%%%%%%%%%%%%%%%%%%%%%%%%%%%%%%%%%%%%%%%%%%%%%%%%%% 
\begin{abstract}
HiCAT is a high-contrast imaging testbed designed to provide complete solutions in wavefront sensing, control and starlight suppression with complex aperture telescopes.  The pupil geometry of such observatories includes primary mirror segmentation, central obstruction, and spider vanes, which make the direct imaging of habitable worlds very challenging. The testbed alignment was completed in the summer of 2014, exceeding specifications with a total wavefront error of 12nm rms over a 18mm pupil. The installation of two deformable mirrors for wavefront control is to be completed in the winter of 2015. In this communication, we report on the first testbed results using a classical Lyot coronagraph. We also present the coronagraph design for HiCAT geometry, based on our recent development of Apodized Pupil Lyot Coronagraph (APLC) with shaped-pupil type optimizations. These new APLC-type solutions using two-dimensional shaped-pupil apodizer render the system quasi-insensitive to jitter and low-order aberrations, while improving the performance in terms of inner working angle, bandpass and contrast over a classical APLC.  
\end{abstract}

%>>>> Include a list of keywords after the abstract 


%%%%%%%%%%%%%%%%%%%%%%%%%%%%%%%%%%%%%%%%%%%%%%%%%%%%%%%%%%%%%
\section{INTRODUCTION}\label{sec:intro}

Several space telescopes are currently investigated (e.g. WFIRST-AFTA\cite{Spergel2015} with monolithic mirror, Large Ultraviolet Optical Infrared [LUVOIR] telescopes with segmented primary mirror, such as ATLAST\cite{Postman2012,Feinberg2014} or HDST\cite{Delcanton2015}) and one of their key goal is the direct imaging and spectroscopy of extrasolar planets, and possibly habitable worlds. However, their aperture geometry makes high-contrast imaging challenging, in particular on two aspects: coronagraphy for starlight suppression and wavefront calibration for contrast stability in the presence of vibrations. We are currently developing HiCAT, a STScI testbed to provide system-level solutions combining wavefront sensing, wavefront control and starlight suppression strategies on such apertures.

Within this framework, we also explore novel high-contrast strategies, including:
\begin{itemize}
\item novel pupil remapping techniques (e.g. Active Control of Aperture Discontinuities [ACAD]\cite{Pueyo2013}) that use two deformable mirrors (DMs) to re-shape complex pupil (including struts and segment gaps) and offer friendly apertures to coronagraph systems. Recent developments are presented in a companion paper by Mazoyer et al. in these proceedings. 
\item innovative coronagraph designs for any type of apertures (e.g. for the Apodized Pupil Lyot Coronagraph [APLC] based on our recent developments for circular axi-symmetric pupil\cite{N'Diaye2015a} or complex geometry apertures\cite{N'Diaye2015b}). 
\end{itemize}
In our previous paper\cite{N'Diaye2014a}, we report on the design overview and the first light results of HiCAT after finalized alignment in the absence of DMs. In this communication, we present our recent developments on HiCAT with the presence of a single facesheet Boston DM and our new findings in coronagraphy for arbitrary apertures.

In Section \ref{sec:HICAT_features}, we present the main features of our testbed. In Section \ref{sec:HiCAT_DMs}, we report on the characterization of our Boston DMs and the integration of a single device in the HiCAT optical train. In Section \ref{sec:HiCAT_WFC}, we present preliminary wavefront control results on the testbed using a single DM and a classical Lyot coronagraph. Finally in Section \ref{sec:coronagraphy}, we show our recent progress in coronagraphy for arbitrary apertures with further implementation in our testbed. 

%%%%%%%%%%%%%%%%%%%%%%%%%%%%%%%%%%%%%%%%%%%%%%%%%%%%%%%%%%%%%
\section{HiCAT main features}\label{sec:HICAT_features}

The testbed optical design was developed using an hybrid approach to define the layout and surface error requirements to minimize amplitude-induced errors, so-called Talbot effects to develop methods for aperture diffraction control. The complete details of our design study are presented in our previous communications\cite{N'Diaye2013b,N'Diaye2014a}. Our testbed uses mirrors with high surface quality (better than $\lambda/40$) based on our specifications that are derived from our high-contrast requirements ($10^{8}$ contrast floor inside a half-field dark hole produced by a single DM with a perfect coronagraph from 3 to 10\,lambda/D with 2\% bandpass). The testbed design includes several features:
\begin{itemize}
\item a 20\,mm pupil mask to mimic the central obstruction and the support struts of the telescope aperture and enable the replication of monolithic telescope geometries such as WFIRST-AFTA.
\item three DMs: an Iris AO segmented mirror to mimic segmentation of the telescope primary mirror with 36 segments, similar to ATLAST/HDST, and 2 Boston kilo-DM for wavefront control and pupil remapping techniques such as ACAD\cite{Pueyo2013}. 
\item a starlight suppression system based on the APLC\cite{Soummer2005} architecture: a reflective apodizer for broadband operations based on the recent developments made by JPL and Princeton\cite{Bala2013} and relying on our recent solutions\cite{N'Diaye2015a}, a reflective, focal plane mask (FPM) inherited from the Lyot project\cite{Oppenheimer2004} and a motorized Lyot stop.
\item two cameras (CamF and CamP) respectively for focal and pupil plane imaging are part of the back end. CamF is positioned on a translation stage to enable direct and coronagraphic phase diversity measurements.\cite{Dean2003,Sauvage2012}
\end{itemize}
The testbed also have additional room for further implementation of low- and mid-order wavefront sensing concepts such as the Zernike wavefront sensors \cite{Wallace2011,N'Diaye2013a}. Additional feature includes the 4D Fizeau interferometer outside the main optical path for alignment purpose and direct wavefront sensing measurements. 

\begin{figure}[!ht]
\centering
\resizebox{0.5\hsize}{!}{
\includegraphics{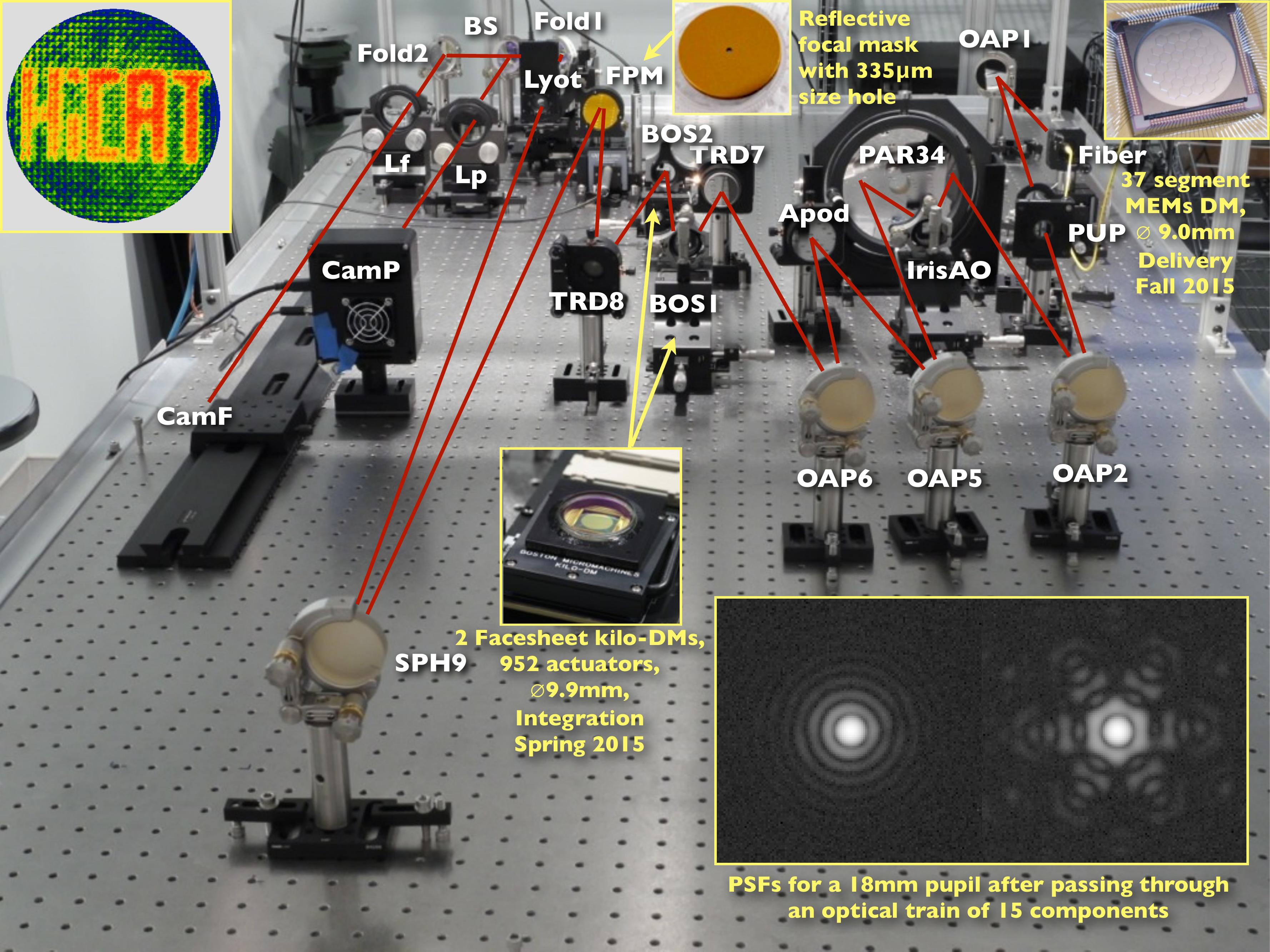}
}
\caption{Overall picture of the HiCAT in June 2014. The alignment was completed in June 2014 in the absence of DMs. We obtained our first PSFs in log scale at the FPM with a 18\,mm unobstructed circular aperture and a JWST-like pupil, see inset images. Seven regular diffraction rings can be observed, confirming the good quality of our alignment and of our optics.}
\label{fig:HiCAT}
\end{figure}

In June 2014, the reflective apodizer and the different DMs were not available yet. They were replaced by flat mirrors with excellent surface error quality ($\sim \lambda$/20 PV surface error over a 2-inch diameter). After procurements of the optical and mechanical parts, we performed the alignment of the testbed and last summer, we obtained our first PSFs in the absence of DMs, see Figure \ref{fig:HiCAT}. We achieved an optical quality of 12nm rms over a 18mm pupil after passing through an optical train of 15 components. At the end, this testbed will combine a segmented DM to mimic the aperture segmentation, a pupil mask for the central obstruction and spiders of the telescope, two Boston DMs for wavefront control and an APLC for starlight suppression. 

%_____________________________________________________________
%_____________________________________________________________

%%%%%%%%%%%%%%%%%%%%%%%%%%%%%%%%%%%%%%%%%%%%%%%%%%%%%%%%%%%%%
\section{HiCAT deformable mirrors}\label{sec:HiCAT_DMs}
\subsection{Environment control}

During last winter, we received our science grade Boston kilo-DMs for wavefront control and pupil remapping applications with HiCAT. These devices must operate in a clean environment with a relative humidity that remains below 30\% for long-term reliability. In particular, humidity is a critical issue since the application of voltage in humid environment leads to corrosion effects on the DM that are irreversible \cite{Shea2004,Morzinski2012}.

\begin{figure}[!ht]
\centering
\resizebox{0.45\hsize}{!}{
\includegraphics{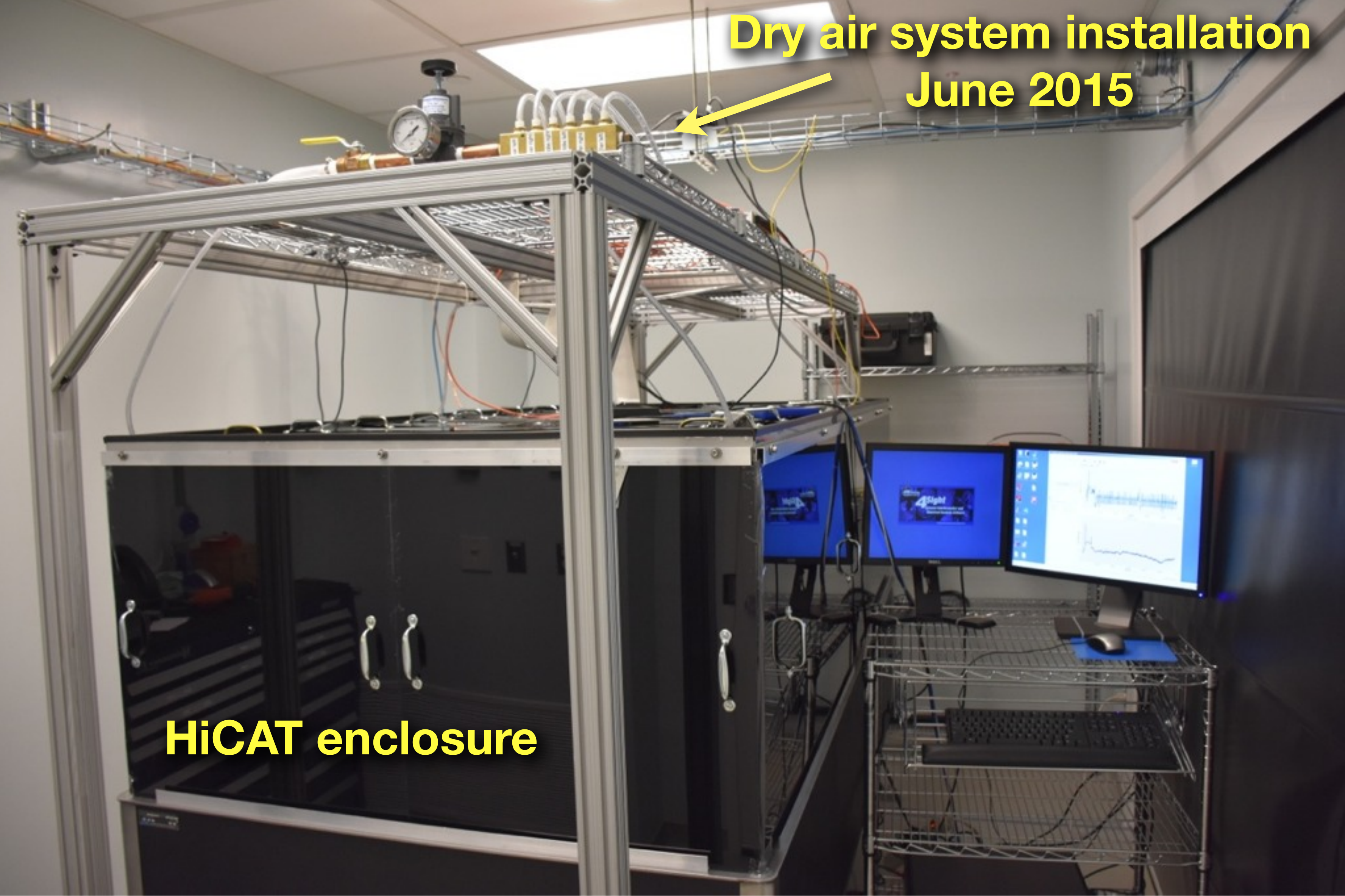}
}
\resizebox{0.45\hsize}{!}{
\includegraphics{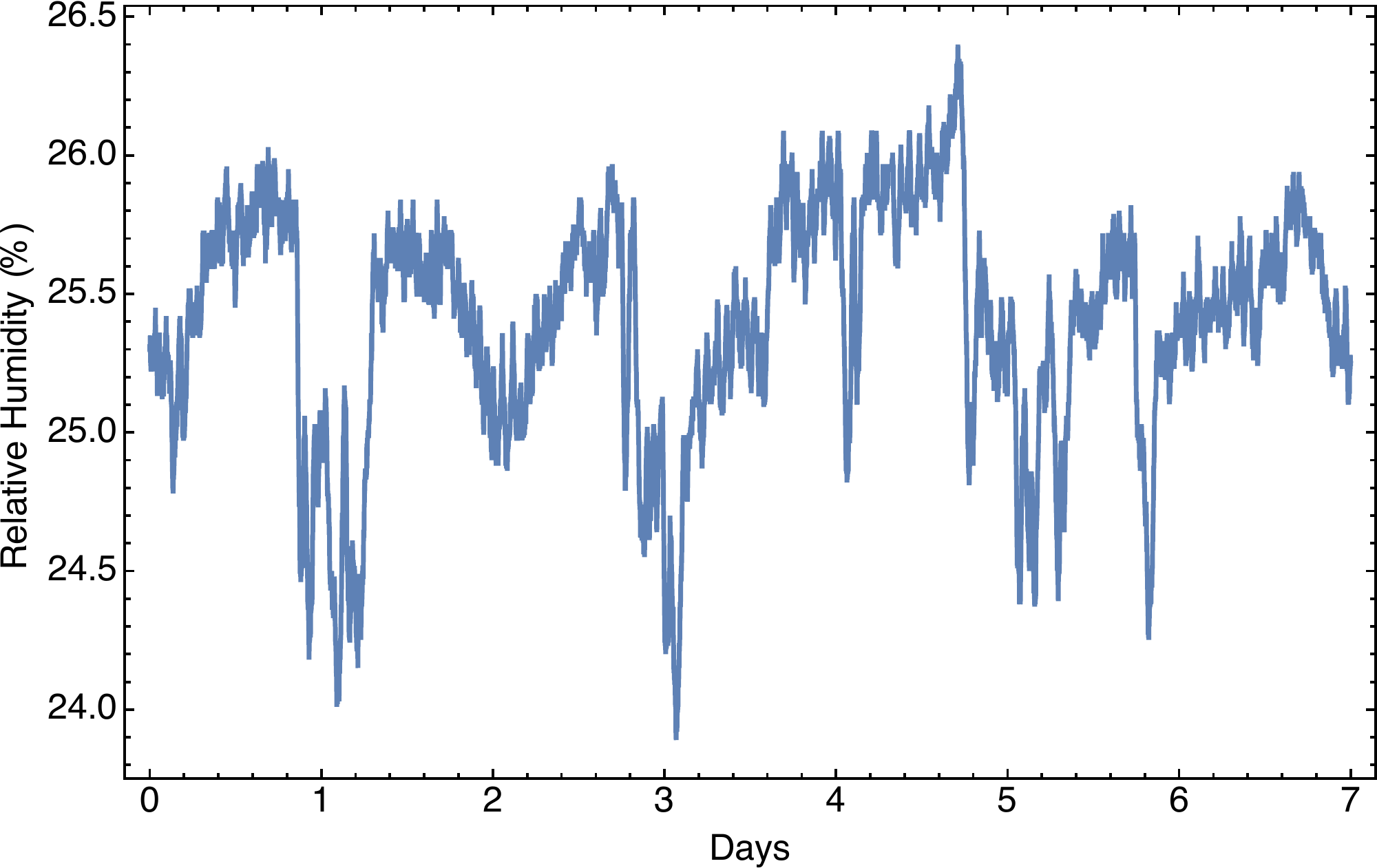}
}
\caption{Environment control for the HiCAT experiment. \textbf{Left}: Picture of the testbed enclosure and its dry air system. The latter was installed in June 2015 in support to the room humidity control system to reach a relative humidity below 30\%, allowing us to operate the DMs without risk of corrosion. \textbf{Right}: Relative humidity measurements obtained during the first week of August 2015, showing relative humidity values around 25\%. Measurements inside the enclosure are permanently performed to monitor the relative humidity and ensure the adequate use of our DMs in the right conditions.}
\label{fig:enclosure_and_dry_air_system}
\end{figure}

HiCAT is located in the STScI Makidon Optics Lab in a class 1000 clean room with a temperature that is controlled to within 1$^{\circ}$C and relative humidity that remains below 40\%. In addition to our lab humidity control, we recently installed a dry air system to our experiment. This system injects 1.5\,psi pressure dry air with the output split between 6 lines to different points of the $\sim$4$m^3$ HiCAT enclosure. With this additional system in our testbed, we keep the relative humidity below 30\%. In addition, we use temperature and humidity sensors to continuously monitor the environment, ensuring the DM is used only in adequate conditions (see Figure \ref{fig:enclosure_and_dry_air_system}).

The dry air system may add unwanted turbulence effects, leading to wavefront variations that makes high-contrast regimes stability challenging. Further plans include the installation of a second stage humidity control within the whole room to keep the humidity below 30\% while minimizing the turbulence effects during experiments.

%_____________________________________________________________
%_____________________________________________________________

\subsection{DM characterization}

Having a good knowledge of our DM behavior is a key point for further wavefront control operations. To obtain such information, we calibrate our DMs using our 4D Fizeau interferometer in front of it. Based on previous analysis studies\cite{Morzinski2012,Mazoyer2014}, we set all the actuators to a DM flat position using a close loop with the 4D Fizeau interferometer. We chose to set this flat position around voltages of 70\% of the maximum voltage of the DM to ensure a maximum actuator stroke range during operation. From this reference point, we apply different voltages values from -60\% to 20\% to every forth actuator across a row and a column. This operation allows us to avoid cross-talk when determining the influence function of each actuators and test all the actuators in just four iterations. Finally, we measure their displacement with the interferometer to determine the actuator influence function. 

\begin{figure}[!ht]
\centering
\resizebox{0.4\hsize}{!}{
\includegraphics{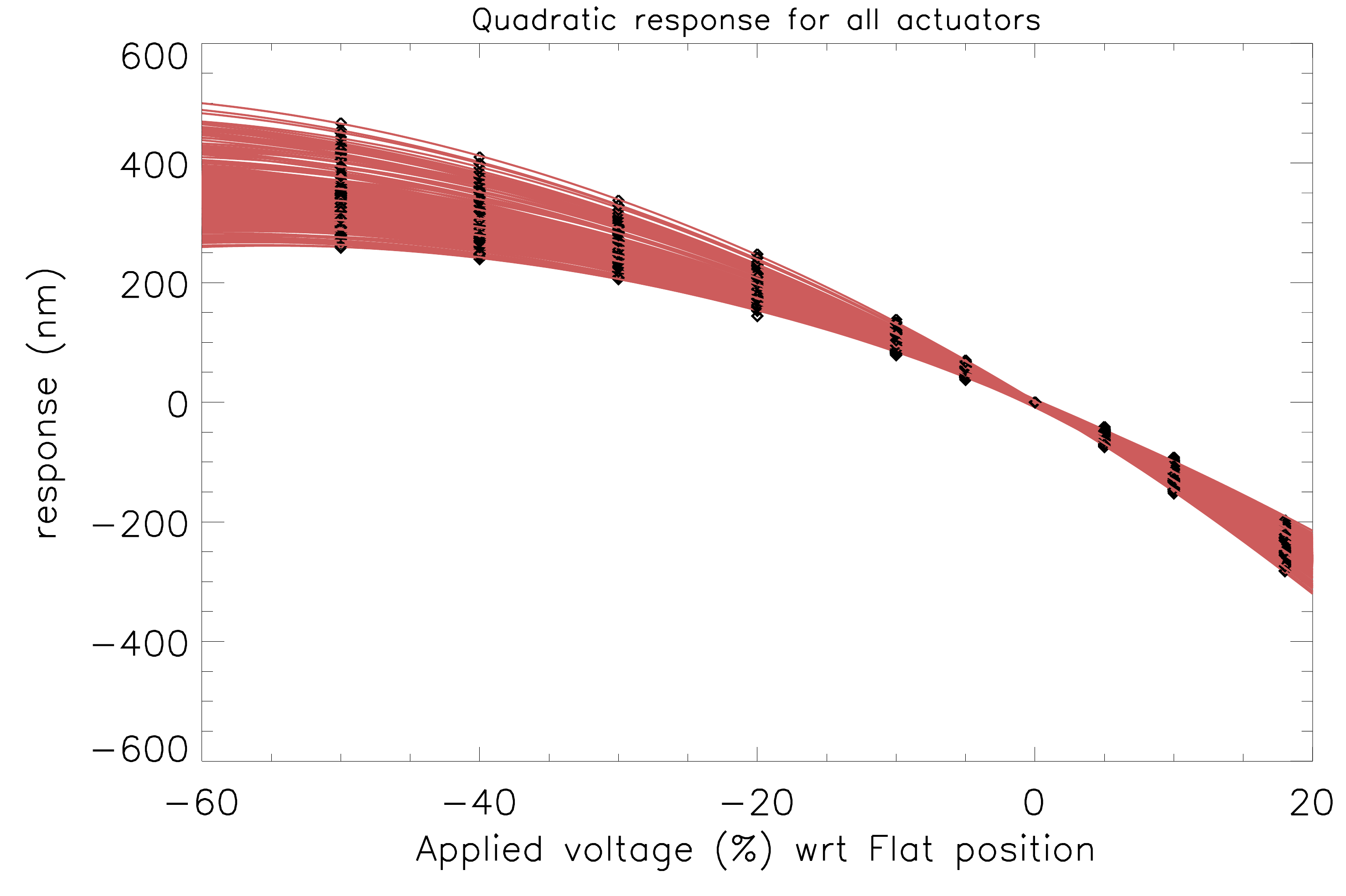}
}
\resizebox{0.5\hsize}{!}{
\includegraphics{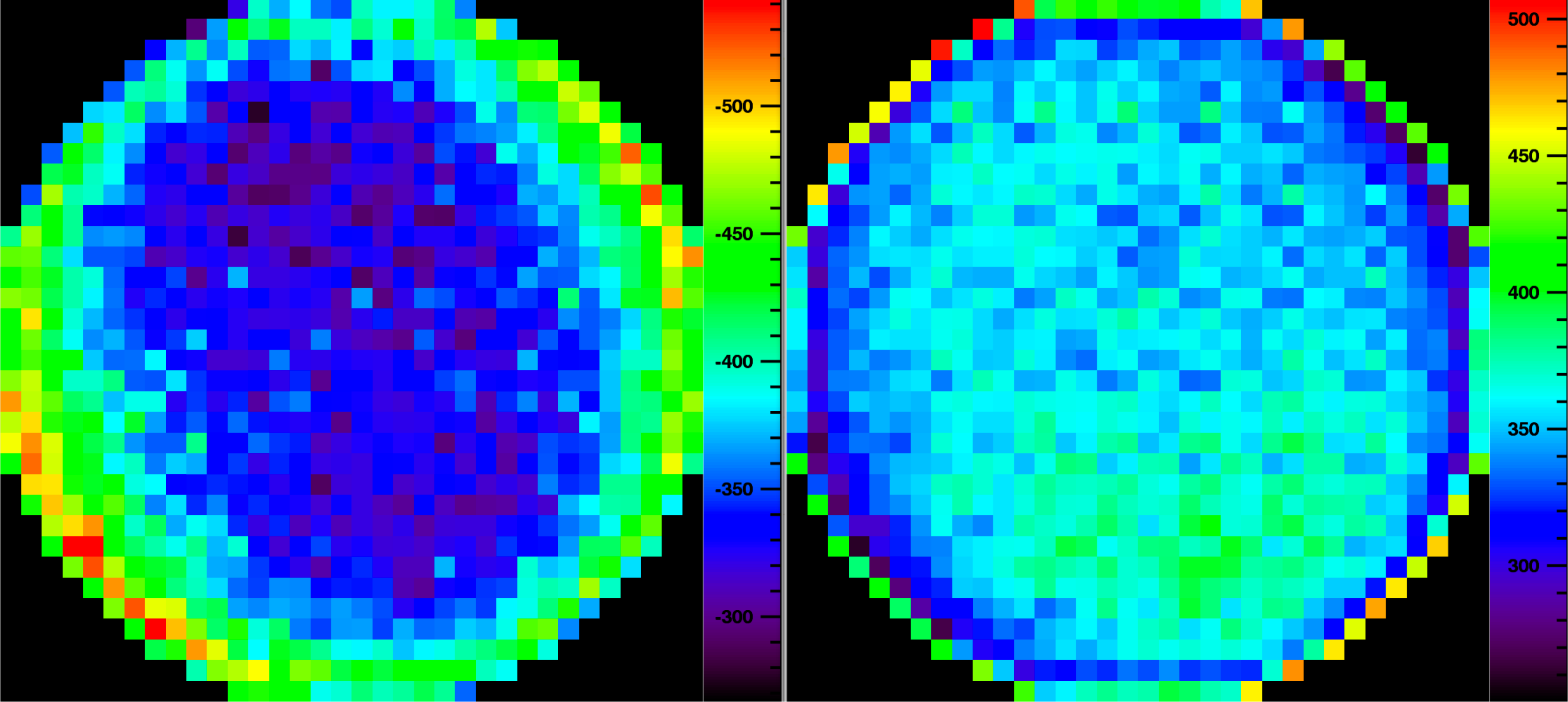}
}
\caption{Characterization of our first Boston DM. \textbf{Left}: Quadratic response curves for each of the 952 actuators of our DM after interpolation of our measurements represented by black dots. For these measurements, we position the DM in front of our Fizeau interferometer, apply a flat reference of the DM (chosen around 70\% of the maximum voltage), and introduce push-pull from this reference to each actuator to determine the influence function for different voltages. The 70\% maximum voltage reference point was chosen to ensure a maximum stroke range for our actuators. After analysis of all our actuators, we estimate a minimum available stroke range of 632\,nm. \textbf{Right}: Maps of the maximum achievable stroke under and above of the flat position for each actuator. As expected, the distribution of strokes follows the chip shape of the DM at rest.}
\label{fig:quadratic_response}
\end{figure}

\begin{figure}[!ht]
\centering
\resizebox{0.4\hsize}{!}{
\includegraphics{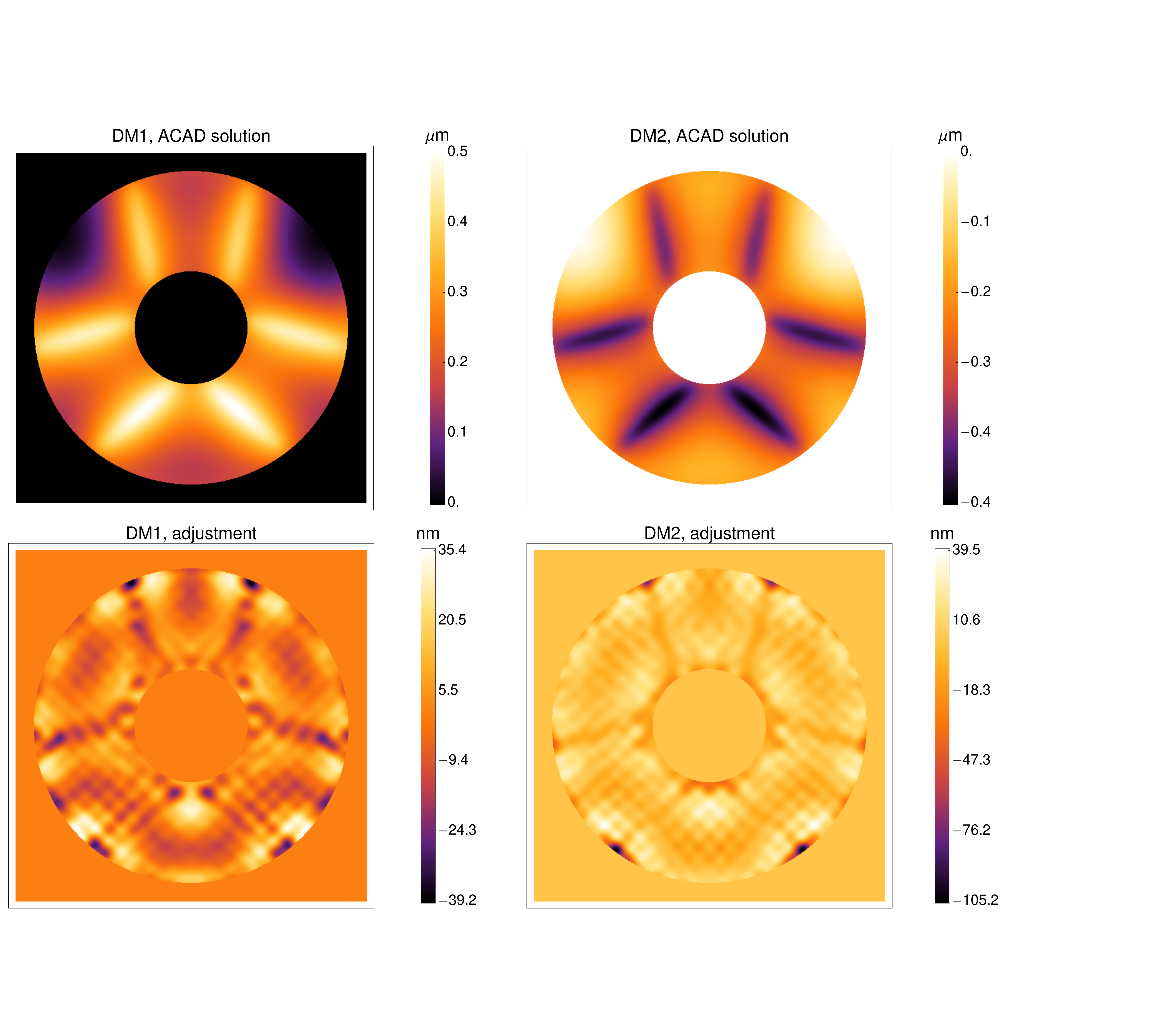}
}
\resizebox{0.55\hsize}{!}{
\includegraphics{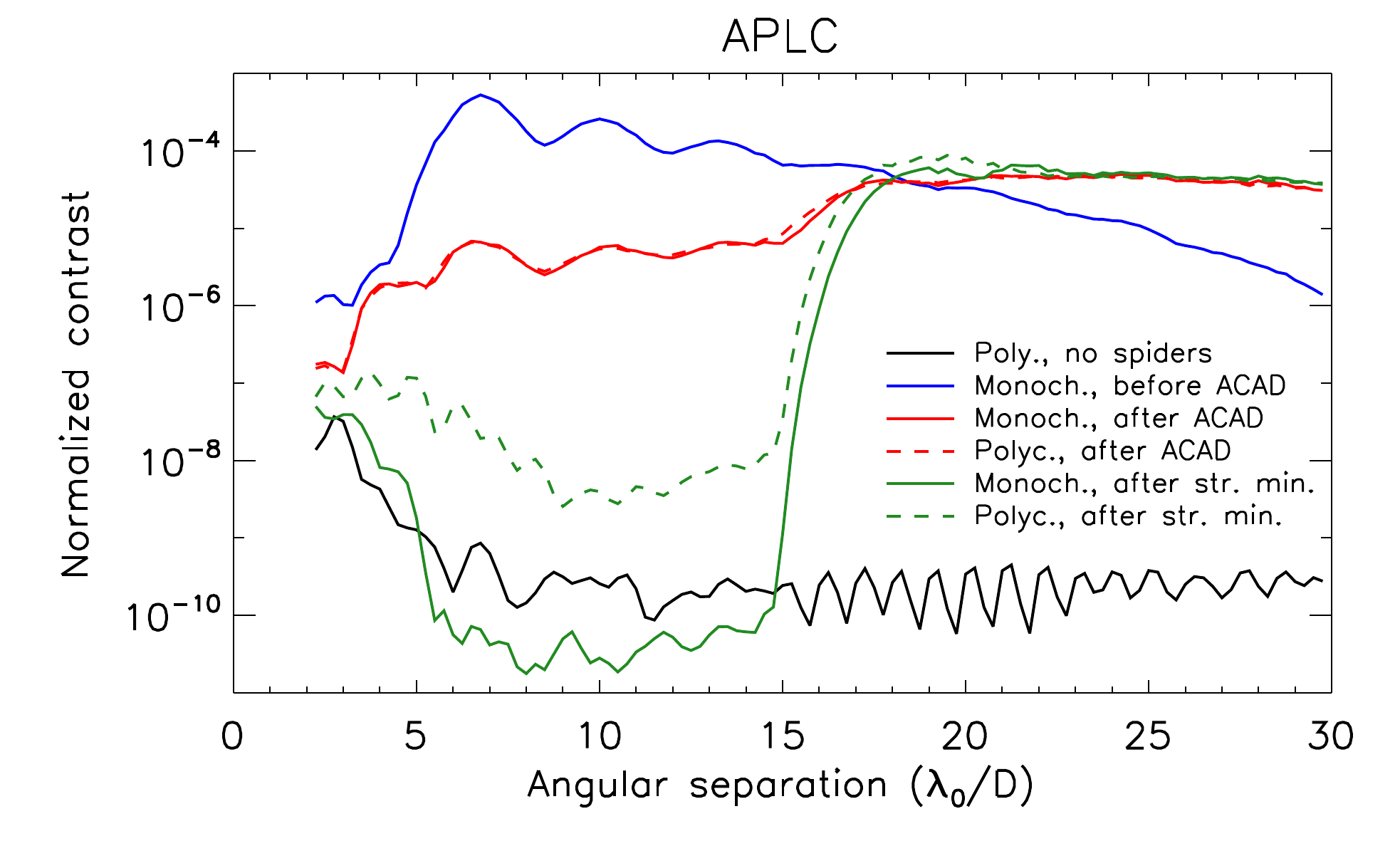}
}
\caption{ACAD studies with WFIRST-AFTA like pupil in preparation for implementation in HICAT. \textbf{Left:} DM shapes at different stages of the ACAD + stroke minimization process for an WFIRST-AFTA like pupil with the HiCAT parameters. Top: Shapes of the ACAD solution for the DM 1 and 2. The deformation strokes are inferior to 450 nm. Bottom: adjustments around this shape after the stroke minimization algorithms. These deformation strokes are inferior to 150 nm. The final strokes (ACAD + Stroke Minimization) are inferior to 470 nm. \textbf{Right:} Radial contrast profile of the images at different stages of the ACAD + stroke minimization process for an WFIRST-AFTA like pupil. These contrasts are normalized by the intensity peak of the PSF without coronagraph. Results in monochromatic and 10\% broadband light are represented in solid and dashed lines. The black curve represents the contrast obtained with an APLC in polychromatic light in the case of a 36\% central obscuration pupil and no struts. The blue line shows the contrast after the introduction of the WFIRST-AFTA support structures of the secondary in the pupil. In red, the contrast in the dark hole after the ACAD solution. Finally, we present the final contrast obtained after ACAD and stroke minimization correction (green curve). A dark hole contrast better than $6.10^{-11}$ is obtained in monochromatic light and $1.10^{-8}$ in broadband light on a 5-15 $\lambda/D$ dark hole, showing the efficiency of our combined solution. From Mazoyer et al. (submitted)\cite{Mazoyer_JATIS2015}.}
\label{fig:ACADsolutionforAFTA}
\end{figure}

We retrieve a quadratic response curve representing the stroke as a function of the applied voltage for each actuator (see Figure \ref{fig:quadratic_response}, left) and determine the maximum, the minimum and the stroke range of every actuators. On Figure \ref{fig:quadratic_response} we show for every actuator the maximum values reachable under (center) and above (right) the reference flat position. From this analysis, we deduce that in the worst case, the minimum stroke range for an actuator around the flat position is 632\,nm. Mazoyer et al. (submitted)\cite{Mazoyer_JATIS2015} recently made a study on the ACAD applicability on the testbed, considering an WFIRST-AFTA-like pupil, our current DMs characteristics and their separation, see Figure \ref{fig:ACADsolutionforAFTA}. The authors estimated a 470\,nm maximum required stroke range to set up the shape of the DMs for application of the pupil remapping technique. This value is well within the measured stroke range for every actuators, confirming the possibility to validate ACAD on HiCAT in laboratory.  

%_____________________________________________________________
%_____________________________________________________________

%_____________________________________________________________
%_____________________________________________________________

\subsection{DM integration}

The Boston DMs have two dedicated locations in the HiCAT optical train, the first one in a pupil plane conjugated to the pupil mask and the second one at 30\,cm away from the first device in an out-of-pupil plane to perform both phase and amplitude error corrections. The science grade DMs were not available during our alignment in June 2014 and we replaced them by high-quality flat mirror ($\lambda/20$ PV surface error over a 2-inch diameter part).

\begin{figure}[!ht]
\centering
\resizebox{0.46\hsize}{!}{
\includegraphics{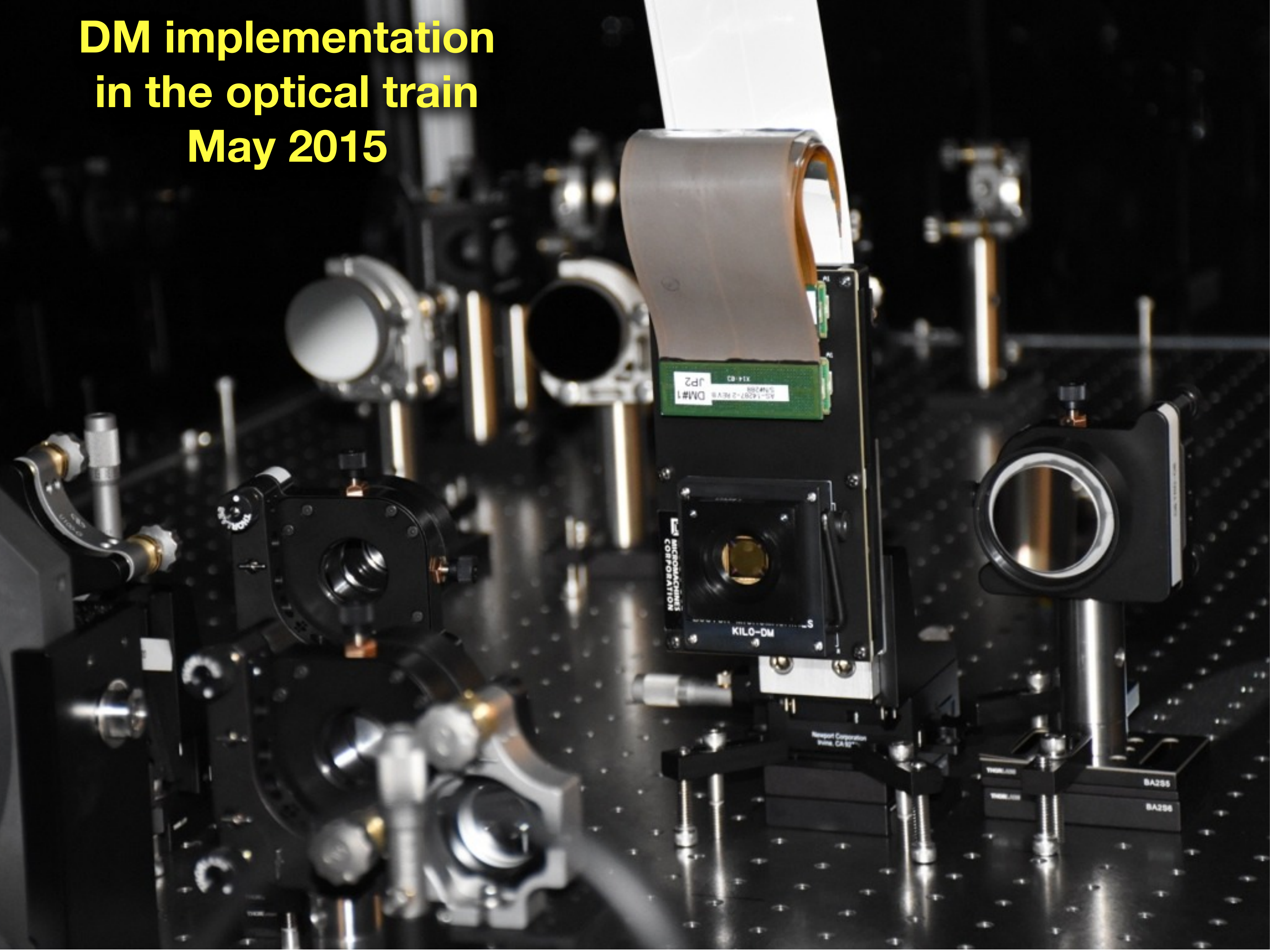}
}
\resizebox{0.44\hsize}{!}{
\includegraphics{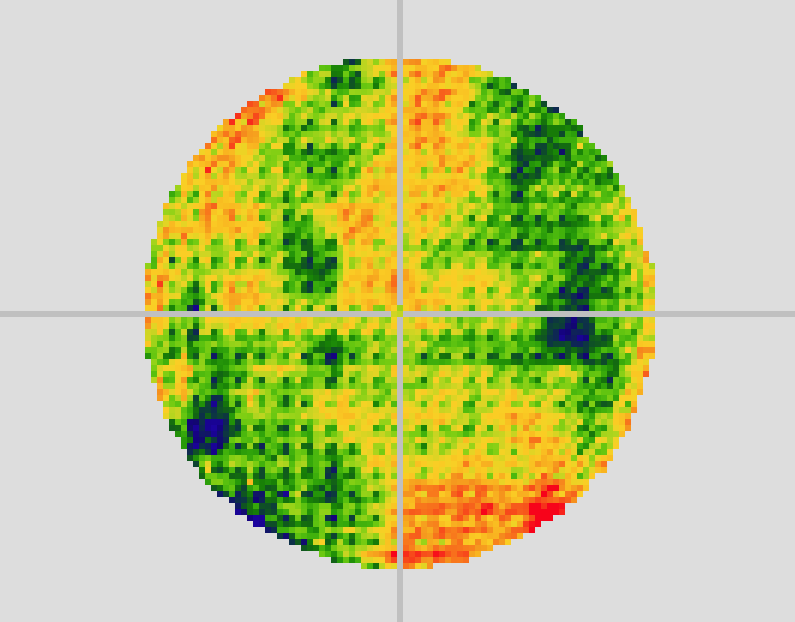}
}
\caption{\textbf{Left:} Picture of the DM after integration in the optical train of the testbed. \textbf{Right}: Example of wavefront map (10.5\,nm rms and 55.0\,nm peak-to-valley wavefront error here) obtained after DM flattening with our Fizeau interferometer after alignment. We measured an optical quality of 13$\pm$3\,nm rms wavefront error after alignment and below 10\,nm rms for the low-order aberrations, reaching a quality quasi-equivalent to the one we obtained before DM insertion. These small values are really promising for the direct application of fine wavefront control algorithms, such as Electric Field Conjugation\cite{Borde2006,Give'on2007} or Stroke Minimization\cite{Pueyo2009}.}
\label{fig:DM_integration}
\end{figure}

After DM characterization, we removed the flat mirror present at the pupil plane location and integrated our first device in Spring 2015. After fine DM positioning, we flattened it and used our interferometer to determine the quality of our alignment. We estimated an optical quality of 13\,$\pm$3\,nm rms wavefront error over 18\,mm size pupil, see Figure \ref{fig:DM_integration}. These values with one DM inserted, with no significant loss in terms of wavefront quality, are almost equivalent to our estimates before insertion, confirming the excellent quality of our DM alignment. 

We also determine a contribution below 10\,nm rms for the low-order aberrations. These small WFE rms values are promising for application of wavefront control algorithms, such as Energy minimization\cite{Borde2006}, Electric Field Conjugation\cite{Give'on2007} or Stroke minimization\cite{Pueyo2009}. Before considering such algorithms and as a first step, we started our work on speckle nulling algorithms to check our system.

%_____________________________________________________________
%_____________________________________________________________

%%%%%%%%%%%%%%%%%%%%%%%%%%%%%%%%%%%%%%%%%%%%%%%%%%%%%%%%%%%%%
\section{First wavefront control results}\label{sec:HiCAT_WFC}

Prior to the 2015 SPIE conference, we started with the implementation of speckle nulling algorithm on HiCAT. Our testbed configuration includes a 18\,mm size circular non obstructed pupil, a single Boston DM, a classical Lyot Coronagraph with 335\,$\mu$m size reflective FPM (5.91\,$\lambda_0/D$ relative size for a F/89 beam at $\lambda_0=638\,$nm) and a 10\,mm diameter Lyot stop. 

We first perform the calibration aspects with our DM in flat position by acquiring coronagraphic images onto CamF. Their analysis allows us to retrieve the plate scale, the optical axis position, the DM orientation and the contrast.

After calibration, we select one speckle and determine the corresponding spatial frequency of the sine wave pattern. We then probe for the speckle phase by first acquiring four images with this sine pattern on the flattened DM with phase shifts of respectively 0, $\pi/2$, $\pi$, $3\pi/2$. We estimate the speckle intensity for each image to interpolate a sine wave function and we derive the phase shift for the sine wave pattern on the flattened DM that minimizes the speckle intensity. We then determine the speckle amplitude based on its relation with the speckle normalized intensity in the initial image.

\begin{figure}[!ht]
\centering
\resizebox{0.45\hsize}{!}{
\includegraphics{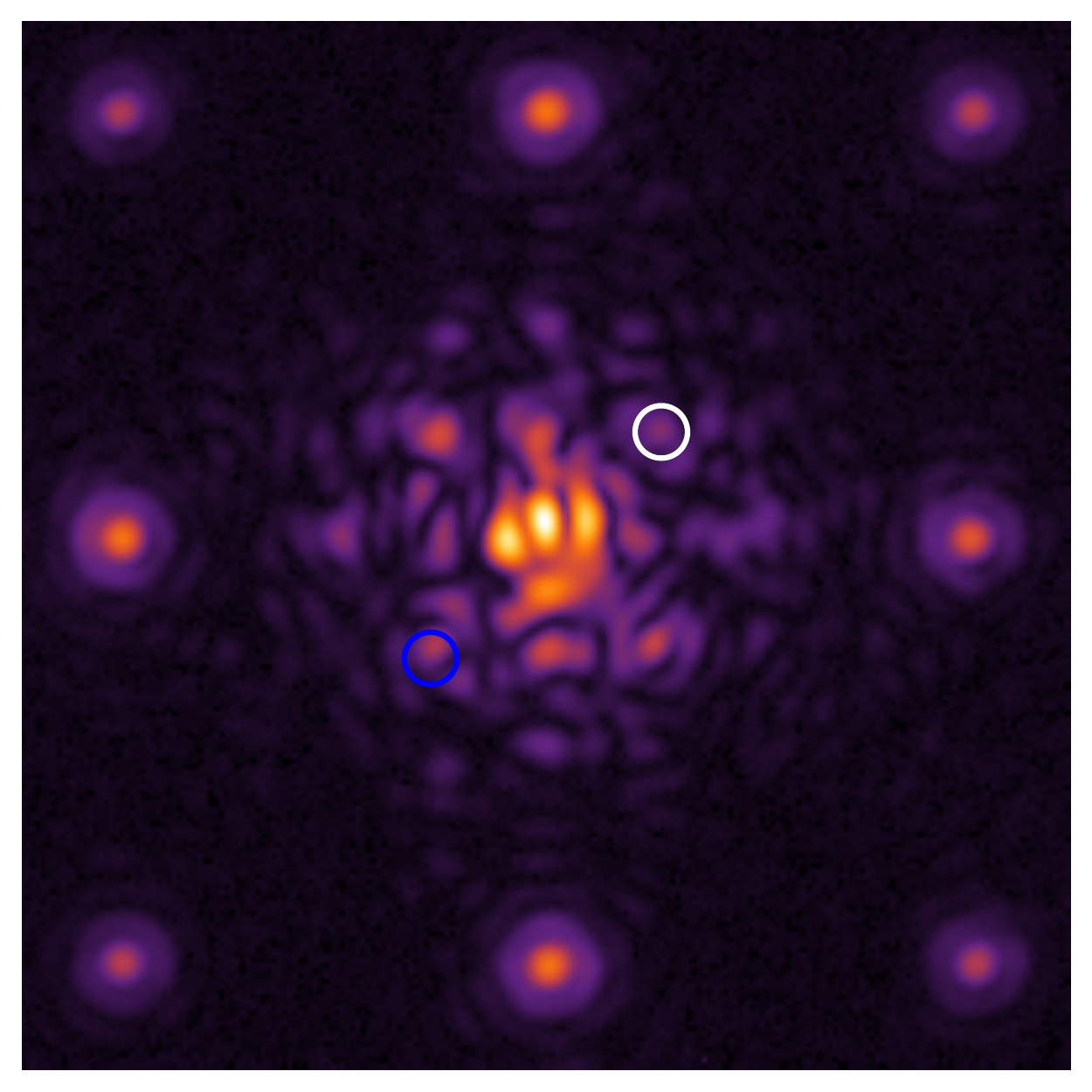}
}
\resizebox{0.45\hsize}{!}{
\includegraphics{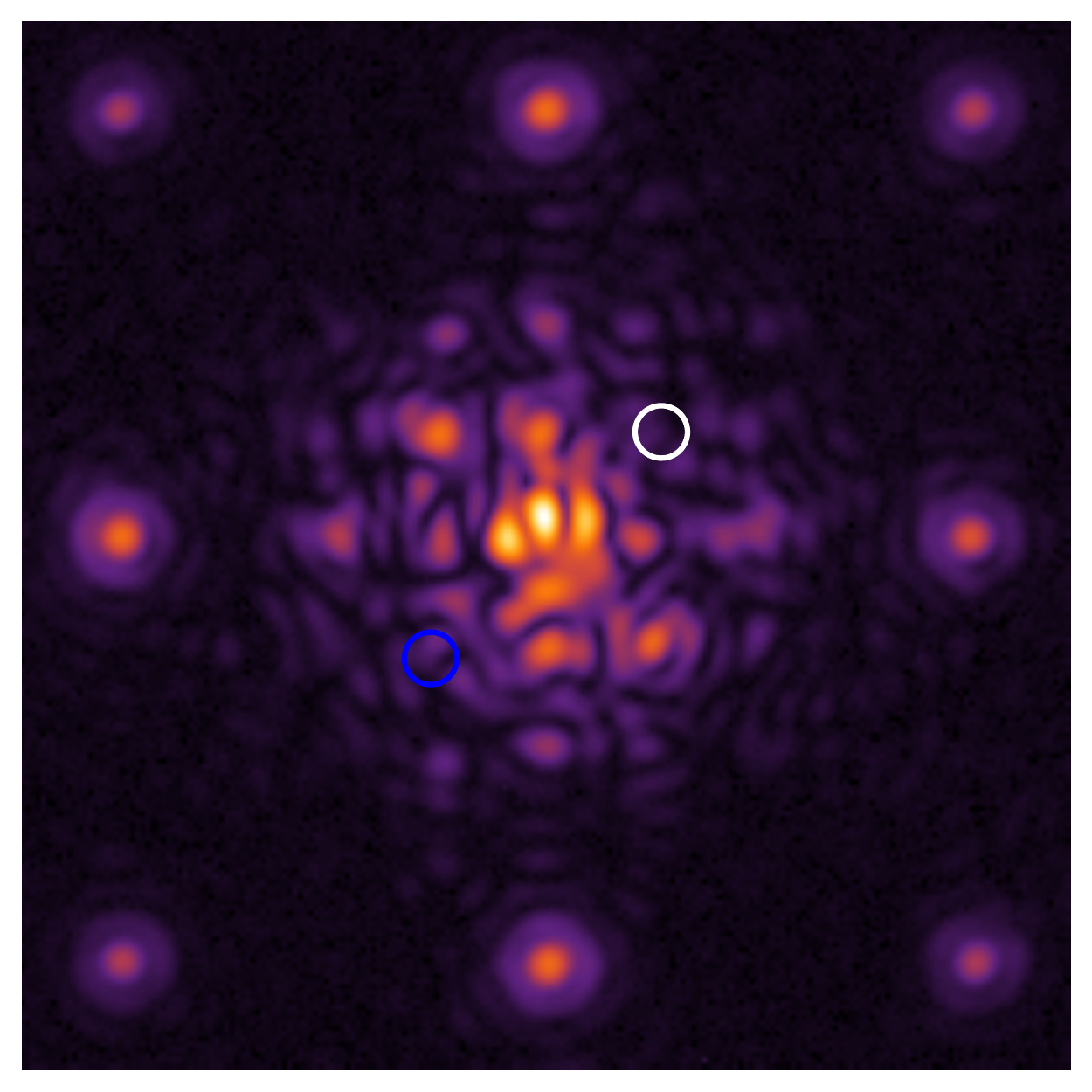}
}
\caption{Coronagraphic images before and after nulling of a single speckle denoted by the white circle. The blue circle represents the symmetric location of the considered speckle with respect to the main optical axis. Light attenuation is observed in both locations, suggesting the domination, at these contrast levels, of phase errors over amplitude errors in HiCAT as we expected when we start the testbed design. Further studies with speckle nulling over a given dark zone will allow us to verify this important aspect for further high-contrast studies with unfriendly apertures.}
\label{fig:Nulling}
\end{figure}

We finally combine the flat position with the estimated sine wave pattern on the DM to attenuate the considered speckle. Figure \ref{fig:Nulling} shows the images before and after DM shaping, showing our first single speckle nulling. The fact that the targeted speckle (circled in white) and its symmetric with respect to the main optical axis (circled in blue) are corrected simultaneously underlines the prevalence of phase errors over amplitude errors at these contrast levels and in that specific image location. This wavefront control step is a preliminary result which paves the way for multi-speckle nulling over a one-side dark hole. 

However, we do not expect high contrast level in this configuration since we are equipped with a classical Lyot coronagraph. In this context and in view of the future exoplanet direct imaging missions with arbitrary apertures, we have been developing novel designs to push starlight suppression further.

%_____________________________________________________________
%_____________________________________________________________

%%%%%%%%%%%%%%%%%%%%%%%%%%%%%%%%%%%%%%%%%%%%%%%%%%%%%%%%%%%%%
\section{Novel coronagraph designs}\label{sec:coronagraphy}

Coronagraphs design for $10^{10}$ contrast performance with large segmented aperture is one of the key issue for the direct imaging and spectroscopy of habitable worlds with future large missions. Several approaches have been studied over the past few months to overcome this issue. Our approach addresses the Apodized Pupil Lyot Coronagraph (APLC), a diffraction suppression system that is currently implemented in the recently deployed exoplanet direct imagers P1640, GPI and SPHERE.

The APLC combines a classical Lyot coronagraph with entrance pupil apodization to achieve starlight suppression \cite{Aime2002, Soummer2003a}. For these instruments, the apodization uses a prolate function which is an eigenfunction of the Lyot-style coronagraphic propagation problem for a given mask size. The designs have been optimized considering the contrast as a metric and exploring the parameter space with these prolate functions and the Lyot stop geometry. These solutions are quasi-achromatic and provides a theoretical contrast of $10^7$ at 0.2" in H-band \cite{Soummer2005,Soummer2011a}. However, these designs are limited in terms of contrast and IWA. 

In this context, we have considered another method to derive novel solutions. This approach is based on shaped pupil type optimization which consists of finding apodizations to produce PSF dark zones. With this method, we increase the number of degrees of freedom with respect to the eigenvalue problem approach: we set the contrast, IWA, bandpass and Lyot stop geometry and we seek for the apodization with the highest throughput that solves the problem. 

\subsection{Circular axi-symmetric pupils}

As a first step, we considered circular axi-symmetric pupils and we found novel solutions that were introduced in N'Diaye et al.\cite{N'Diaye2015a}. In the context of GPI, we showed our ability to gain one order of magnitude in terms of contrast and 1\,$\lambda$/D in IWA\cite{N'Diaye2015a}.

More interestingly, we showed the existence of solutions which produces PSF core smaller than the projected coronagraphic mask. This allows PSF movement within the mask with no impact on the coronagraph performance. Our new concepts are therefore more robust to tip, tilts, focus drifts and other vibrations, a major concern in coronagraphy. 

In the context of GPI, we show that our new designs offer a better contrast and a lower sensitivity to low-order aberrations. We also provide $10^{10}$ contrast designs with an IWA that depends on the bandpass and the central obstruction size, see N'Diaye et al. for further details\cite{N'Diaye2015a}.

\subsection{Arbitrary apertures}

\subsubsection{Solution for large segmented aperture}

These proposed solutions essentially apply to circular axi-symmetric apertures, corresponding to a one dimension radial problem. The presence of spiders struts and segment gaps in the pupil will alter the coronagraph performance. This effect can be mitigated with the use of ACAD solutions\cite{Pueyo2013} upstream the diffraction suppression system.

An alternative, complementary approach consists of developing coronagraphic solutions that work with unfriendly pupils. Both Princeton University and STScI teams recently develop novel approaches in the context of WFIRST-AFTA and ATLAST based on APLC and shaped pupil coronagraphs. These approaches extend our previous method for one-dimension radial problem to two-dimension problem. The Princeton team develops the Shaped pupil approach and combines it with a classical Lyot coronagraph to increase the coronagraph performance. At STScI, we consider the APLC and look for apodizers with Shaped pupil solutions. These close paths were studied in parallel in the context of WFIRST by Princeton and in the context of segmented apertures by STScI. 

\begin{figure}[!ht]
\centering
\resizebox{\hsize}{!}{
\includegraphics{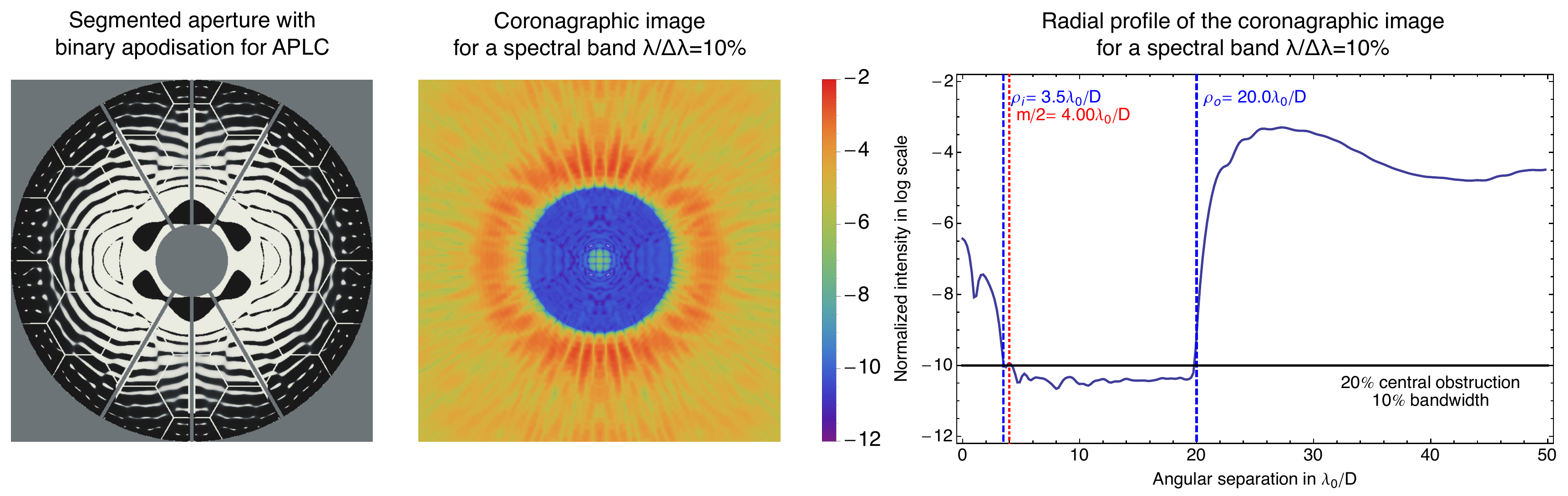}
}
\caption{\textbf{Left}: Shaped pupil apodization for the APLC coronagraph for the ATLAST mission. \textbf{Middle}: Coronagraphic image obtained with the coronagraph for a 10\% spectral bandwidth. A dark zone with a $10^{10}$ contrast level is produced with the segmented aperture, paving the way for the observation of habitable worlds. \textbf{Right}: Radial intensity profile of the previous image with the size of the focal plane mask and the dimensions of the high-contrast region delimited in red and blue. The stellar image core is smaller than the mask, allowing an enlargement or a displacement within the mask without impact on the contrast zone, making the design virtually insensitive to low-order aberrations, such as pointing errors or defocus drifts \cite{N'Diaye2015b}.}
\label{fig:atlast_corono_design}
\end{figure}

We consider the telescope and coronagraph parameters given in Table \ref{table:params} and following our developed approach, we recently found solutions with APLC using a shaped pupil that provides a $10^{10}$ contrast PSF dark zone over a 10\% bandpass, with a moderate inner working angle, see Figure \ref{fig:atlast_corono_design}. 

\subsubsection{Apodizer manufacturing aspects}

Our solution uses an 600-point diameter across apodization that is not fully binary. Increasing the number of points results in a heavy computational problems that cannot be resolved with our current resources. We consider another approach starting from our current numerical solution. We first test a design with rounded values of our numerical solution. This leads to a contrast loss of two orders of magnitude. 

\begin{figure}[!ht]
\centering
\resizebox{0.5\hsize}{!}{
\includegraphics{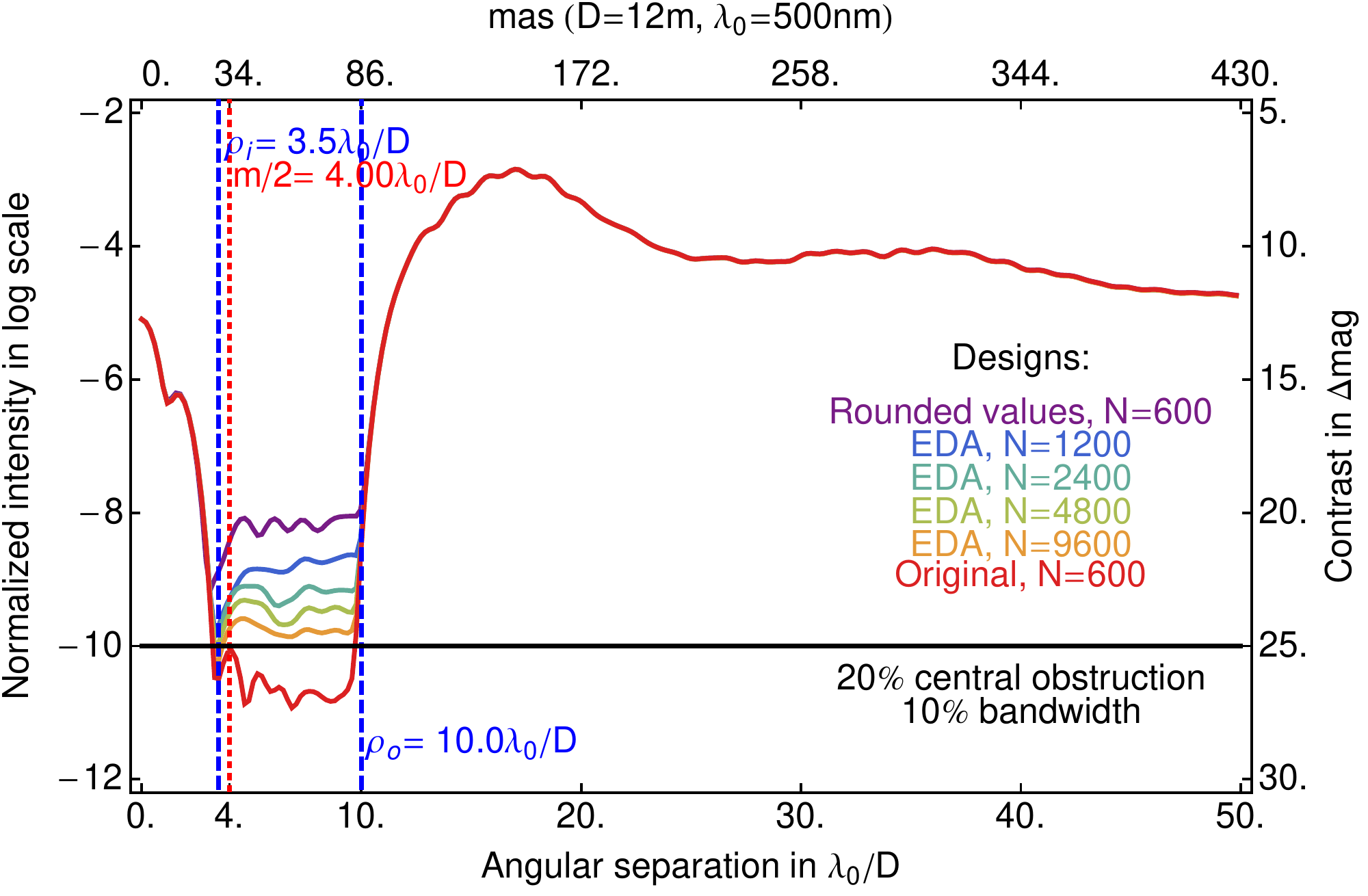}
}
\caption{Azimuth averaged intensity profile of the coronagraphic image reached by an APLC solution in 10\% broadband light and the parameters shown in Table \ref{table:params}, except for the Lyot stop which presents a 36\% instead of 40\% obstruction, providing a quasi-binary shaped pupil. Profiles are represented for the design with the original grey version, rounded values version, and the binarized versions using error diffusion algorithm (EDA, see e.g. \cite{Dorrer2007}) with different lateral size $N$ obtained by sub-pixelization of the original design gray pixels. The dashed blue vertical lines delimit the high-contrast search area $\mathcal{D}$ ($\rho_i=3.5\,\lambda_0/D$ and $\rho_o=$10.0\,$\lambda_0/D$). The red dot line delimits the FPM radius, set to $m/2=4\,\lambda_0/D$. The averaged contrast over the spectral band in the dark region is below $10^{-10}$ (black horizontal line). The $10^{10}$ contrast performance of the original design is almost recovered with a EDA version using N=9600. Relating on current black silicon technologies to manufacture Shaped pupil mask for WFIRST-AFTA coronagraph \cite{Bala2013}, we translate these values into physical units. Assuming a 15\,$\mu$m size for a pixel, the apodizer of our design can currently be fabricated with a 144\,mm diameter prototype to work in visible light.}
\label{fig:design_binarity}
\end{figure}

To recover the ultimate $10^{10}$ contrast, we use an binary approximation which combines a sub-pixellisation of the current solution points and the error diffusion algorithm \cite{Dorrer2007}. With a 16 factor pixellization, we achieved a binary design that almost provides a $10^{10}$ contrast, see Figure \ref{fig:design_binarity}.

Based on current technologies for the fabrication of a reflective apodizer,  we assume a black silicon pattern mask with 15\,$\mu$m size pixel. A binary mask for our design is feasible with a 144\,mm diameter prototype. This is a very encouraging result since our design is manufacturable as-of-today with current technologies and further improvements on size reduction or design are expected to reduce the prototype size. 

This concept is furthermore robust to low-order aberrations, providing room for small vibrations of the instrument. Working with large telescope however leads to observation of resolved stars and stellar angular size is a major concern in coronagraphy\cite{Guyon2006}. 

%_____________________________________________________________
%_____________________________________________________________

%_____________________________________________________________
%_____________________________________________________________

%_____________________________________________________________
\begin{table}[!ht]
\caption{Parameters for the design showed in Figure \ref{fig:atlast_corono_design}.}
\centering
\begin{tabular}{c c}
\hline\hline
parameters & value\\
\hline
Contrast C & 10\\
Dark zone inner edge radius $\rho_i$ & 3.5\,$\lambda_0/D$\\
Dark zone outer edge radius $\rho_o$ & 20.0\,$\lambda_0/D$\\
focal plane mask radius $m/2$ & 4.0\,$\lambda_0/D$\\\hline
Aperture & 20\% central obstruction\\
& 1\% aperure size spiders\\
& 0.2\% aperture size gaps\\\hline
Lyot stop & 40\% central obstrcution\\
& 2\% aperutre size spiders\\
& no segmentation\\\hline
Broadband optimization & 3 wavelengths within 10\% band\\
\hline
\end{tabular}\\
\label{table:params}
\end{table}
%_____________________________________________________________

\subsubsection{Sensitivity to stellar angular size}

\begin{figure}[!ht]
\centering
\resizebox{0.5\hsize}{!}{
\includegraphics{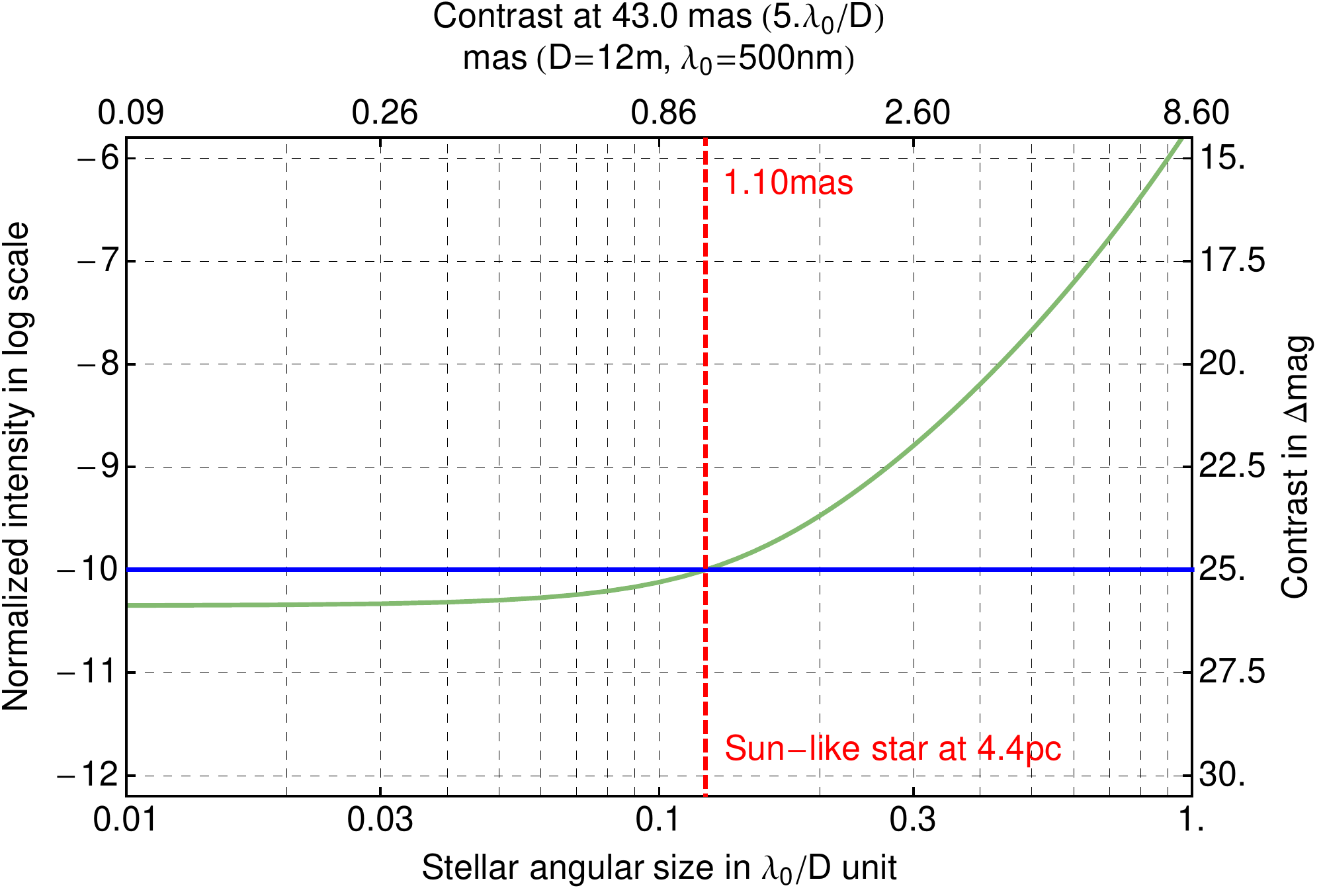}
}
\caption{Averaged intensity of the broadband coronagraphic image at a 5\,$\lambda_0/D$ angular separation from the optical axis as a function of the stellar angular size. The curve is obtained for our design with $\rho_1=$20\,$\lambda_0/D$. Blue solid line denotes the $10^{-10}$ intensity level. Our design presents a plateau and intensity levels below $10^{-10}$ for stellar angular size up to 0.12 $\lambda_0/D$, underlining the contrast performance stability of our coronagraph design. Assuming a 12\,m telescope at 500\,nm, our coronagraph is robust to stellar angular sizes up to 1.1\,mas, allowing observations of planets around Sun-like star located beyond 4.3\,pc.}
\label{fig:atlast_star_size}
\end{figure}

We study the impact of stellar leaks on our concept due to stellar angular size. Figure \ref{fig:atlast_star_size} show the contrast at 5\,$\lambda_0$/D as a function of the stellar angular size for a 12\,m telescope at 500nm. We can notice a plateau on the contrast curve, showing the stability of our concept to resolved stars. Actually, our concept can observe a Sun-like star beyond 4.4\,pc with a 12\,m telescope without being impacted by the stellar angular size, proving very encouraging for future direct imaging space missions.

%_____________________________________________________________
%_____________________________________________________________

\begin{figure}[!ht]
\centering
\resizebox{0.35\hsize}{!}{
\includegraphics{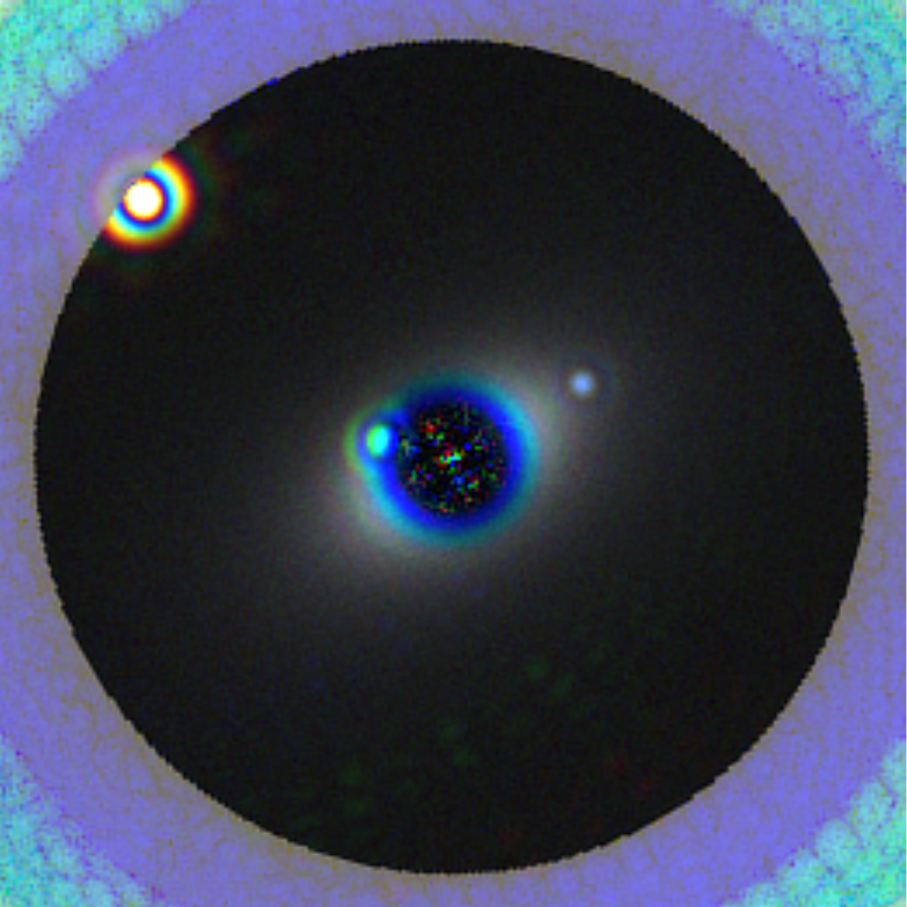}
}
\caption{Simulated multicolor pictures of a solar system twin at 13.5 pc with a 12 m telescope and our coronagraph design with an OWA $\rho_1=$30\,$\lambda_0$/D at different exposure times (40h). The solar system was modeled with the Haystacks project (See A. Roberge et al., http://asd.gsfc.nasa.gov/projects/haystacks/haystacks.html), including spectroscopic and spatial information for all the components within the exoplanetary system. Each picture is a composite of three images on three channels around 400, 500 and 600 nm, each with 10\% bandpass and a coronagraph optimized for this band. In this simulation, we assume a perfect wavefront, no wavefront drifts between the target "solar system" star and calibrator star for the image processing, and only photon noise. Earth is at 2 o'clock and is indeed blue, Venus is at 9 o'clock, Zodiacal light is elongated along the 2-8 o'clock direction, Jupiter is at 10 o'clock in the red channel (600 nm): it is outside of the dark hole at shorter wavelengths. In that channel, most of Venus is hidden by the coronagraph. Linear and logarithmic scale representations are used inside and outside the high-contrast region for each coronagraph.}
\label{fig:earth-like_planet_image}
\end{figure}

Following these results, we decide to simulate the image of a solar system twin with our concept. We use a model based on spectral and spatial information from Haystack project and based on these assumptions, we produce the image in Figure \ref{fig:earth-like_planet_image} in which an Earth-like planet can be observed and characterize after 40h exposure time.

%_____________________________________________________________
\section{Conclusions}
HiCAT is an operating testbed that enables on-axis monolithic or segmented aperture coronagraph system-level studies. The STScI facility will allow us to develop integrated solutions for unfriendly geometry aperture such as WFIRST-AFTA or HDST. To perform such studies, our  testbed will rely on novel diffraction control and wavefront control schemes with multiple DMs (one segmented Iris AO and 2 Boston Micromachines).

This year, we went from the aligned testbed without DMs to preliminary results in wavefront control on testbed with one integrated DM. Our studies include the characterization of Boston DM, its integration in the HiCAT optical train, and the first single speckle nulling with classical Lyot coronagraph.
 
We have also explored novel coronagraph designs for arbitrary telescope apertures with further implementation in the HiCAT testbed for lab validation. We found designs that provide a $10^{10}$ contrast at 4\,$\lambda_0$/D with moderate 18\% Airy throughput for a large segmented aperture with our hybrid APLC/Shaped Pupil solution, representing a first proof of existence. Our solutions prove quasi-insensitive to low-order aberrations and stellar angular size. In addition, these solutions are manufacturable with as-of-today technologies and are fully functioning. 

Our next steps includes the generation of a one side dark hole using speckle nulling algorithm during fall. We will then consider the introduction of the second Boston DM for application of ACAD in laboratory. Based on our recent studies, we will design apodizer prototypes for starlight suppression with different aperture geometries. In the long term, we will have three DMs and we will set novel system level solutions including diffraction suppression and wavefront control for high-contrast generation and stability with arbitrary apertures, advancing technologies for direct imaging and spectroscopy of habitable worlds with future space missions.

%%%%%%%%%%%%%%%%%%%%%%%%%%%%%%%%%%%%%%%%%%%%%%%%%%%%%%%%%%%%%
\acknowledgments     %>>>> equivalent to \section*{ACKNOWLEDGMENTS}       
This work is supported by the National Aeronautics and Space Administration under Grants NNX12AG05G and NNX14AD33G issued through the Astrophysics Research and Analysis (APRA) program (PI: R. Soummer). This material is also partially based upon work carried out under subcontract 1496556 with the Jet Propulsion Laboratory funded by NASA and administered by the California Institute of Technology. The authors warmly acknowledge Tyler Groff, N. Jeremy Kasdin, Charles-Philippe Lajoie, Bruce Macintosh, Dimitri Mawet, Colin Norman, J. K. Wallace, and Stuart Shaklan for fruitful discussions during the testbed design. The authors are also very grateful to STScI and its staff members, in particular Bill Franz, Joe Hunkeler and Kelly Coleman, for their invaluable support. %The authors are also grateful to Nicole Cade-Ferreira, Andrew Frazier, Matthew Jorgensen, and Tucker Kearney from the Department of the Mechanical Engineering of Johns Hopkins University, for the design, development and integration of the HiCAT testbed enclosure.

%L.P., M.P.D., and R.S. conceived of this project and led the overall effort. M.N. and E.C. developed the whole design and simulation studies for the testbed. O.L. realized the opto-mechanical design. S.E. implemented the optical and mechanical design, and with L.L., they performed the fine alignment of the testbed and capture images. E.E. helped with the optical design and the tolerancing studies. A.D. performed the drawings of the optics for the manufacturers. J.K.W. provided valuable advice and guidance for the alignment. C.L. helped with assembly and machining. E.H., M.M. and M.F. developed and manufactured the 6-inch parabola and the toric mirrors and they also contributed to the optical design. As lab manager R.A. coordinated support and lab infrastructure for the new testbed. 

%%%%%%%%%%%%%%%%%%%%%%%%%%%%%%%%%%%%%%%%%%%%%%%%%%%%%%%%%%%%%
%%%%% References %%%%%
\bibliography{biblio_mam}   %>>>> bibliography data in report.bib
\bibliographystyle{spiebib}   %>>>> makes bibtex use spiebib.bst

\end{document}